\documentclass[11pt,letterpaper]{article}

\usepackage[margin=1in]{geometry}
\usepackage{amsmath,amssymb,amsthm}
\usepackage{booktabs}
\usepackage{array}
\usepackage{tabularx}
\usepackage{xcolor}
\PassOptionsToPackage{hyphens}{url}  %% allow URL line-breaks at hyphens
\usepackage{hyperref}            %% load before natbib and microtype
\usepackage{natbib}              %% provides \citep, \citet
\usepackage[expansion=false]{microtype}  %% load after hyperref to avoid patch warning
\usepackage{setspace}
\usepackage{titlesec}
\usepackage{enumitem}
\usepackage{graphicx}
\usepackage{caption}
\usepackage{float}
\usepackage{tikz}
\usetikzlibrary{shapes.geometric,arrows.meta,positioning,calc,decorations.pathreplacing,fit,backgrounds}
\usepackage{mdframed}
\usepackage{tcolorbox}
\tcbuselibrary{theorems,breakable,skins}

\definecolor{TealDark}{HTML}{01696F}
\definecolor{TealMid}{HTML}{0C8E96}
\definecolor{TealLight}{HTML}{BCE2E7}
\definecolor{NavyText}{HTML}{1F2937}
\definecolor{MutedGray}{HTML}{6B7280}
\definecolor{RuleLine}{HTML}{D1D5DB}
\definecolor{TheoremBg}{HTML}{F0FAFA}
\definecolor{DefBg}{HTML}{F7F6F2}
\definecolor{WarnAmber}{HTML}{92400E}

\hypersetup{
  colorlinks=true,
  linkcolor=TealDark,
  citecolor=TealDark,
  urlcolor=TealMid,
  pdftitle={The Theorems of Dr.\ David Blackwell and Their Contributions to Artificial Intelligence: A Survey},
  pdfauthor={Napoleon Paxton},
  pdfsubject={Survey, Statistics, Machine Learning, Artificial Intelligence},
  pdfkeywords={Blackwell, Rao-Blackwell, Approachability, Informativeness, Reinforcement Learning, RLHF, SLAM, Variance Reduction}
}

\titleformat{\section}{\large\bfseries\color{TealDark}}{\thesection.}{0.6em}{}[\vspace{-2pt}\color{RuleLine}\hrule\vspace{4pt}]
\titleformat{\subsection}{\normalsize\bfseries\color{NavyText}}{\thesubsection}{0.6em}{}
\titleformat{\subsubsection}{\normalsize\itshape\bfseries\color{NavyText}}{\thesubsubsection}{0.6em}{}
\titlespacing*{\section}{0pt}{14pt}{4pt}
\titlespacing*{\subsection}{0pt}{10pt}{2pt}
\titlespacing*{\subsubsection}{0pt}{8pt}{1pt}

\tcbset{
  theoremstyle/.style={
    enhanced,
    breakable,
    colback=TheoremBg,
    colframe=TealDark,
    fonttitle=\bfseries\color{white},
    coltitle=white,
    attach boxed title to top left={yshift=-2mm,xshift=4mm},
    boxed title style={colback=TealDark,colframe=TealDark,sharp corners},
    sharp corners=south,
    left=6pt, right=6pt, top=8pt, bottom=6pt,
    before upper={\setlength{\parskip}{3pt}},
  },
  defstyle/.style={
    enhanced,
    breakable,
    colback=DefBg,
    colframe=MutedGray,
    fonttitle=\bfseries\color{NavyText},
    coltitle=NavyText,
    attach boxed title to top left={yshift=-2mm,xshift=4mm},
    boxed title style={colback=DefBg,colframe=MutedGray,sharp corners},
    sharp corners=south,
    left=6pt, right=6pt, top=8pt, bottom=6pt,
    before upper={\setlength{\parskip}{3pt}},
  }
}

\newtcbtheorem[number within=section]{blackwellthm}{Theorem}{theoremstyle}{thm}
\newtcbtheorem[number within=section]{blackwelldef}{Definition}{defstyle}{def}
\newtcbtheorem[number within=section]{blackwellcor}{Corollary}{theoremstyle}{cor}

\newenvironment{abstractbox}{%
  \begin{mdframed}[
    linecolor=TealDark,linewidth=0.8pt,
    backgroundcolor=TheoremBg,
    leftmargin=0pt,rightmargin=0pt,
    innerleftmargin=12pt,innerrightmargin=12pt,
    innertopmargin=10pt,innerbottommargin=10pt,
    skipabove=10pt,skipbelow=10pt
  ]
  \small
}{%
  \end{mdframed}
}

\begin{document}

%% ── Title page ──────────────────────────────────────────────────────────────
\begin{center}
  {\LARGE\bfseries\color{TealDark}
    The Theorems of Dr.\ David Blackwell\\[4pt]
    and Their Contributions to Artificial Intelligence\\[2pt]
    \large\mdseries\color{NavyText} A Survey}\\[10pt]
  {\large Napoleon Paxton}\\[2pt]
  {\normalsize School of Information, University of California, Berkeley}\\[2pt]
  \color{RuleLine}\hrule
\end{center}

\vspace{4pt}

\begin{abstractbox}
\textbf{Abstract.}
Dr.\ David Blackwell (1919--2010) was a mathematician and statistician of the first rank,
whose contributions to statistical theory, game theory, and decision theory predated many of
the algorithmic breakthroughs that define modern artificial intelligence.
This survey examines three of his most consequential theoretical results---the
\emph{Rao-Blackwell theorem}, the \emph{Blackwell Approachability theorem}, and the
\emph{Blackwell Informativeness theorem} (comparison of experiments)---and traces their direct
influence on contemporary AI and machine learning.
We show that these results, developed primarily in the 1940s and 1950s, remain technically
live across modern subfields including Markov Chain Monte Carlo inference, autonomous mobile
robot navigation (SLAM), generative model training, no-regret online learning, reinforcement
learning from human feedback (RLHF), large language model alignment, and information design.
NVIDIA's 2024 decision to name their flagship GPU architecture ``Blackwell'' provides vivid
testament to his enduring relevance.
We also document an emerging frontier: explicit Rao-Blackwellized variance reduction in LLM
RLHF pipelines, recently proposed but not yet standard practice.
Together, Blackwell's theorems form a unified framework addressing information compression,
sequential decision-making under uncertainty, and the comparison of information
sources---precisely the problems at the core of modern AI.
\end{abstractbox}

\vspace{6pt}
\noindent\textbf{Keywords:} Rao-Blackwell theorem; Blackwell approachability; comparison of experiments; sufficient statistics; variance reduction; no-regret online learning; reinforcement learning from human feedback; SLAM; information design; survey.

\noindent\textbf{MSC 2020:} 62B05 (sufficiency and information); 62C05 (Bayes procedures and decision theory); 91A15 (Stochastic games, stochastic differential games); 62-02 (Research exposition).

\noindent\textbf{ACM CCS 2012:} Computing methodologies~$\to$~Machine learning; Theory of computation~$\to$~Online learning theory; Theory of computation~$\to$~Algorithmic game theory.
\tableofcontents
\newpage

%% ════════════════════════════════════════════════════════════════════════════
\section{Introduction}
%% ════════════════════════════════════════════════════════════════════════════

The convergence of statistics and artificial intelligence is one of the defining intellectual
stories of the late twentieth century. While AI research in its early decades was dominated by
symbolic reasoning and rule-based systems, the shift toward probabilistic and data-driven
methods that began in the 1980s and accelerated through the 1990s brought statistical theory
from the periphery to the center of the field. Nowhere is this shift more dramatically
illustrated than in the work of David Harold Blackwell, whose theoretical contributions from
the 1940s and 1950s anticipated problems that would not become computationally tractable for
decades.

Blackwell worked at the intersection of statistics, game theory, and decision theory at a time
when these fields were just being formalized. His collaborators included John von Neumann,
Leonard Savage, Kenneth Arrow, and Richard Bellman---the architects of much of modern
mathematical economics and optimization theory. Yet Blackwell's own contributions, while
equally foundational, have received less recognition outside specialist circles.
This survey aims to correct that by tracing the influence of his three principal theorems
through modern AI.

In March 2024, NVIDIA unveiled its ``Blackwell'' GPU architecture---a 208-billion-transistor
chip designed explicitly for the generative AI era. The naming was a recognition that
Blackwell's statistical and game-theoretic frameworks laid the conceptual groundwork for the
computational advances that make large-scale AI possible today. This paper examines the
specific theoretical mechanisms by which that groundwork was laid.

\subsection{Related Work}

Several adjacent survey literatures inform the present work.
For reinforcement learning from human feedback (RLHF), the most comprehensive surveys are
\citet{kaufmann2023survey}, which covers feedback types, reward modeling, policy learning, and
theory, and \citet{wirth2017survey}, which provides the foundational treatment of
preference-based RL predating LLM-era RLHF.
\citet{casper2023open} survey open problems and alignment challenges in RLHF.
For online learning and regret minimization---the arena in which Blackwell Approachability has
had its deepest modern impact---\citet{cesabianchi2006prediction} provide the canonical
textbook treatment, while \citet{hoi2021online} survey applied online learning algorithms.
\citet{lattimore2020bandit} unify bandit algorithms and explicitly trace connections to
approachability. For Markov chain Monte Carlo (MCMC) methods, \citet{andrieu2003introduction}
provide the widely cited survey-style tutorial bridging MCMC theory and ML practice.
Blackwell's own 1965 work on discounted dynamic programming \citep{blackwell1965discounted}
predates all of these and is recognized in recent economics-oriented RL surveys
\citep{rawat2024survey} as foundational to the Bellman--Howard--Blackwell dynamic programming
framework.

No prior survey covers all three of Blackwell's theorems (Rao-Blackwell, Approachability,
Informativeness) jointly in the context of AI and machine learning. Works citing Blackwell
typically treat one theorem in isolation. This survey's contribution is to draw these lines
together and demonstrate that the three results form a coherent and unified foundation for the
information-processing challenges at the core of modern AI.

\subsection{Summary: Blackwell's Theorems in AI}

Table~\ref{tab:summary} maps each theorem to its principal AI application areas and
representative papers. Figure~\ref{fig:triangle} provides a visual synthesis.

\begin{table}[h!]
\centering
\caption{Blackwell's theorems and their AI applications.}
\label{tab:summary}
\renewcommand{\arraystretch}{1.25}
\begin{tabularx}{\textwidth}{>{\bfseries}l c X l}
\toprule
Theorem & Year & AI Subfield & Representative Paper \\
\midrule
Rao-Blackwell   & 1947 & MCMC Inference                        & \citet{liu1994covariance} \\
Rao-Blackwell   & 1947 & SLAM / Indoor AMR Navigation           & \citet{doucet2001rao} \\
Rao-Blackwell   & 1947 & Generative Model Training              & \citet{liu2019rao} \\
Rao-Blackwell   & 1947 & Policy Gradient / RLHF (emerging)      & \citet{tucker2017rebar}; \citet{zhu2025kl} \\
Approachability & 1956 & No-Regret Online Learning              & \citet{abernethy2011blackwell} \\
Approachability & 1956 & Calibrated Forecasting                 & \citet{foster1998asymptotic} \\
Approachability & 1956 & Multi-Objective RLHF                   & \citet{xiong2025multiobjective} \\
Approachability & 1956 & Fair Online Learning                   & \citet{chzhen2021unified} \\
Informativeness & 1951 & Information Design                     & \citet{bergemann2019information} \\
Informativeness & 1951 & AI Alignment / Safety                  & \citet{alignmentforum2023blackwell} \\
Informativeness & 1951 & Active Learning                        & \citet{settles2009active} \\
Discounted DP   & 1965 & Reinforcement Learning                 & \citet{sutton2018reinforcement} \\
\bottomrule
\end{tabularx}

\vspace{3pt}
{\footnotesize Abbreviations used in this table: AMR, autonomous mobile robot; DP, dynamic
programming; MCMC, Markov chain Monte Carlo; RLHF, reinforcement learning from human feedback;
SLAM, simultaneous localization and mapping.}
\end{table}

%% ════════════════════════════════════════════════════════════════════════════
\section{Biographical Context}
%% ════════════════════════════════════════════════════════════════════════════

David Harold Blackwell was born on April 24, 1919, in Centralia, Illinois. A prodigious
talent, he completed his PhD at the University of Illinois at Urbana-Champaign in 1941 at age
22, under the supervision of Joseph Doob, with a dissertation on Markov chains. He was
awarded a fellowship at the Institute for Advanced Study in Princeton, where he encountered
John von Neumann, whose work on game theory would profoundly influence Blackwell's subsequent
research.

After positions at Southern University and Clark Atlanta University, Blackwell joined Howard
University in 1944, where he rose to chair the mathematics department. In 1954, he moved to
the University of California, Berkeley, where he would spend the remainder of his career and
chair the Statistics Department from 1957 to 1961. In 1965, he became the first African
American elected to the National Academy of Sciences. He was later elected to the American
Academy of Arts and Sciences (1968) and received the John von Neumann Theory Prize in 1979.

Blackwell published more than 90 papers and supervised 64 doctoral dissertations. His
monograph co-authored with M.A.\ Girshick, \emph{Theory of Games and Statistical Decisions}
\citep{blackwell1954theory}, became a foundational text in statistical decision theory.
His collaborators included Leonid Hurwicz, Kenneth Arrow, Richard Bellman, and Lester Dubins.
He passed away on July 8, 2010, having shaped multiple generations of statisticians,
economists, and---though he could not have known it at the time---the AI researchers who
would follow.

%% ════════════════════════════════════════════════════════════════════════════
\section{The Rao-Blackwell Theorem}
%% ════════════════════════════════════════════════════════════════════════════

This section states the Rao-Blackwell theorem and then traces it through four applications in
which conditioning on a sufficient statistic reduces the variance of a Monte-Carlo estimate:
MCMC, particle-filter SLAM, generative-model training, and policy-gradient reinforcement
learning. Of the three theorems surveyed here it is the most directly operational---working
implementations invoke it by name---and it is therefore treated first.

\subsection{Theorem Statement and Significance}

The Rao-Blackwell theorem was established independently by C.\,R.\ Rao in 1945
\citep{rao1945information} and David Blackwell in 1947 \citep{blackwell1947conditional};
the name honours both contributions equally. We first establish the necessary notion of
sufficiency.

\begin{blackwelldef}{Sufficient Statistic}{suff}
Let $X$ be data with distribution $P_\theta$ indexed by parameter $\theta$.
A statistic $T = T(X)$ is \emph{sufficient} for $\theta$ if the conditional distribution
$P_\theta(X \mid T = t)$ does not depend on $\theta$ for any value $t$.
Equivalently (Fisher--Neyman factorization), the likelihood factors as
$f_\theta(x) = g_\theta(T(x))\,h(x)$,
so $T$ captures all information about $\theta$ contained in $X$.
\end{blackwelldef}

\begin{blackwellthm}{Rao-Blackwell Theorem (Blackwell, 1947)}{rb}
Let $S = S(X)$ be an unbiased estimator of $\theta$ (i.e.\ $\mathbb{E}_\theta[S] = \theta$
for all $\theta$), and let $T$ be a sufficient statistic for $\theta$.
Define the \emph{Rao-Blackwellized estimator}
\[
  S^* = \mathbb{E}[S \mid T].
\]
Then:
\begin{enumerate}[noitemsep,label=(\roman*)]
  \item $S^*$ is unbiased: $\mathbb{E}_\theta[S^*] = \theta$;
  \item $S^*$ has weakly lower variance: $\operatorname{Var}(S^*) \leq \operatorname{Var}(S)$,
        with equality if and only if $S$ is already a function of $T$ almost surely.
\end{enumerate}
\end{blackwellthm}

The proof of part (ii) follows immediately from the law of total variance:
$\operatorname{Var}(S) = \operatorname{Var}(\mathbb{E}[S|T]) + \mathbb{E}[\operatorname{Var}(S|T)]
= \operatorname{Var}(S^*) + \mathbb{E}[\operatorname{Var}(S|T)] \geq \operatorname{Var}(S^*)$.
The process of computing $S^* = \mathbb{E}[S \mid T]$ is called \emph{Rao-Blackwellization}.
It provides a constructive recipe---not merely an existence result---for improving any
unbiased estimator. The theorem is a cornerstone of mathematical statistics, underpinning
the theory of minimum variance unbiased estimators (UMVUE).

To illustrate: if one estimates the probability $\theta$ that a coin lands heads using only
the first of $n$ flips ($S = X_1$), the sufficient statistic is $T = \sum_{i=1}^n X_i$.
Rao-Blackwellization gives $S^* = T/n$, the sample mean, reducing variance from
$\theta(1-\theta)$ to $\theta(1-\theta)/n$.

\subsection{Applications in AI and Machine Learning}

\subsubsection{MCMC Variance Reduction}

Markov Chain Monte Carlo methods are the workhorses of Bayesian inference, enabling posterior
computation in models where analytical solutions are unavailable---including Bayesian neural
networks, probabilistic graphical models, and hierarchical models widely used in AI.
A central challenge is that sample-based estimates of posterior expectations carry substantial
variance, particularly in high-dimensional models.
Rao-Blackwellization provides a principled remedy: rather than computing $f(X_i)$ for each
MCMC sample, one replaces it with $\mathbb{E}[f(X_i) \mid \mathcal{G}_i]$, where
$\mathcal{G}_i$ is a sigma-algebra generated by a subset of the variables.
\citet{liu1994covariance} demonstrated that this extra conditioning consistently improves
Monte Carlo estimates---a result that has become standard in the MCMC toolkit.
In practice, this yields smoother posterior estimates for the same computational budget,
directly improving the reliability of Bayesian AI systems.

\subsubsection{Rao-Blackwellized Particle Filters and the Indoor Robotics Revolution}
\label{sec:rbpf}

Simultaneous Localization and Mapping (SLAM)---the problem of building a map of an unknown
environment while simultaneously tracking one's position within it---is a foundational
capability for autonomous mobile robots (AMRs) operating in dynamic indoor environments such
as warehouses, factories, hospitals, and retail distribution centers. The computational
challenge is severe: the joint state space of robot pose and environment map is enormous, and
naive particle filters require prohibitively many particles.

\citet{doucet2001rao} introduced the Rao-Blackwellized Particle Filter (RBPF), which factors
the problem: the robot's pose trajectory is sampled via particles (sequential Monte Carlo),
while the map conditioned on each pose is maintained analytically via closed-form Kalman
filter updates. Since map features are conditionally linear-Gaussian given the trajectory,
their posterior can be computed in closed form---this is precisely the Rao-Blackwell step
that replaces high-variance sampling with a lower-variance analytical estimate.

There is a deeper theoretical connection here: the Kalman filter subroutine inside RBPF is
itself an instance of Blackwell's 1965 discounted dynamic programming framework
\citep{blackwell1965discounted}. The Kalman filter solves a linear-Gaussian optimal
estimation problem that can be cast as a dynamic programming recursion over a sequence of
observation steps; Blackwell's existence and uniqueness results for stationary optimal
policies in discounted infinite-horizon MDPs provide the theoretical guarantee that this
recursion converges to the unique optimal estimator. The RBPF thus simultaneously instantiates
two of Blackwell's contributions---the 1947 Rao-Blackwell theorem at the particle-filter
level and the 1965 DP theorem at the Kalman-filter level.

The GMapping system \citep{grisetti2007improved} implemented RBPF-SLAM and became the
standard in the Robot Operating System (ROS) ecosystem, deployed across warehouse, logistics,
and service robot platforms worldwide.

The commercial relevance of this work has intensified dramatically. While outdoor autonomous
vehicles have adopted sensor-fusion and deep-learning navigation stacks that go beyond
classical RBPF, the dominant growth wave in robotics is now \emph{indoor} autonomous mobile
robots---the domain where RBPF-based SLAM remains highly competitive. The technical reason is
specific: RBPF-SLAM handles the noisy, cluttered, GPS-denied environments of real warehouses
and factory floors, where the satellite positioning that outdoor stacks lean on is unavailable.
The Rao-Blackwell theorem, via RBPF, therefore sits at the algorithmic core of a large and
still-growing class of deployed indoor robots.

Independent market forecasts consistently project double-digit compound growth for this sector,
driven by e-commerce, labor shortages, and demand for same-day fulfilment. Those figures
describe the commercial context in which RBPF-SLAM operates; they are not evidence for any
mathematical claim made in this survey, and they are collected separately in
Appendix~\ref{app:market} so that the distinction is unambiguous.

\subsubsection{Variance Reduction in Generative Model Training}

Training variational autoencoders (VAEs) and other latent-variable generative models requires
estimating gradients of an Evidence Lower Bound (ELBO) via Monte Carlo sampling. When latent
variables are discrete---as in models generating structured text, categorical codes, or
symbolic representations---standard reparameterization tricks are unavailable, and
REINFORCE-style gradient estimators exhibit notoriously high variance.
Rao-Blackwellization addresses this directly.
\citet{liu2019rao} introduced Rao-Blackwellized stochastic gradient estimators for discrete
distributions, replacing the raw REINFORCE gradient with its conditional expectation over a
subset of the sampled variables.
\citet{paulus2020rao} applied the same idea to the Gumbel-Softmax trick, producing a
Rao-Blackwellized straight-through estimator with substantially lower variance.
In both cases the underlying principle is Blackwell's: conditioning on available information
to produce lower-variance estimates.

\subsubsection{Policy Gradient Variance Reduction: Classical RL and the LLM Frontier}

The policy-gradient variance reduction story has two distinct chapters: a well-established
classical chapter covering general RL and discrete latent-variable models, and an emerging
chapter where explicit Rao-Blackwellization is just beginning to enter large language model
(LLM) training pipelines.

\paragraph{Classical RL and discrete models.}
In the policy-gradient setting, Rao-Blackwellization means replacing an unbiased gradient
estimator $g(X, Y)$ with its conditional expectation $\mathbb{E}[g(X, Y) \mid X]$,
analytically marginalizing over some of the stochasticity.
This provably reduces variance without bias.
\citet{liu2019rao} formalized this for categorical distributions, and \citet{tucker2017rebar}
introduced REBAR---a Rao-Blackwellized baseline using a continuous relaxation of discrete
variables---substantially reducing gradient variance in discrete latent-variable models.
\citet{ranganath2014black} established the broader black-box variational inference framework
that preceded these explicit Rao-Blackwellized estimators.
These methods are well-validated in academic benchmarks and remain the most direct application
of the Rao-Blackwell theorem to RL.

\paragraph{LLM RLHF pipelines: an emerging frontier.}
Mainstream LLM fine-tuning via RLHF---as used in alignment based on Proximal Policy
Optimization (PPO) and Trust Region Policy Optimization (TRPO)---applies variance reduction
primarily through classical RL devices: reward-to-go, value-function baselines, advantage
functions, and Generalized Advantage Estimation (GAE). These are control-variate
methods in spirit, but are not typically framed or implemented as explicit
Rao-Blackwellizations. Open-source RLHF libraries generally estimate sequence-level KL
penalties and their gradients with straightforward Monte Carlo estimators.
\citet{zhu2025kl} make this gap explicit: they derive a Rao-Blackwellized estimator for the
sequence-level KL between a policy LM and a reference LM---conditioning on token prefixes and
analytically summing over continuations---and note that this estimator is ``absent from
existing literature and open-source RLHF libraries.''
Their experiments demonstrate substantially lower variance, more stable RLHF training, and
policies appearing more frequently on the reward-KL Pareto frontier.
This constitutes strong evidence that explicit Rao-Blackwellized variance reduction in LLM
RLHF is \emph{just beginning} to be developed, not yet a codified standard.
Practical adoption barriers include vocabulary-scale compute cost, adequacy of existing
baselines, and the ecosystem inertia of large production RLHF stacks.

%% ════════════════════════════════════════════════════════════════════════════
\section{The Blackwell Approachability Theorem}
%% ════════════════════════════════════════════════════════════════════════════

This section states the Approachability theorem and follows it into no-regret online learning,
calibrated forecasting, reinforcement learning for MDPs, multi-objective alignment, and fair
online learning. Where Rao-Blackwell governs estimation from a fixed sample, approachability
governs repeated decisions against an adversary, and its central object is a vector of payoffs
rather than a scalar loss. The tightness of the connection to current AI practice varies
considerably across the applications below, and each is qualified where it is not exact.

\subsection{Theorem Statement and Significance}

\begin{blackwelldef}{Vector-Payoff Repeated Game}{vpgame}
Consider a two-player repeated game where, at each round $t$, Player~1 chooses action
$a_t \in \mathcal{A}$, Player~2 chooses action $b_t \in \mathcal{B}$, and the outcome is a
vector payoff $r_t = r(a_t, b_t) \in \mathbb{R}^d$.
The \emph{time-averaged payoff} after $T$ rounds is
$\bar{r}_T = \tfrac{1}{T}\sum_{t=1}^T r_t$.
A closed convex set $\mathcal{S} \subset \mathbb{R}^d$ is \emph{approachable} by Player~1
if Player~1 has a strategy guaranteeing
$d(\bar{r}_T, \mathcal{S}) \to 0$ almost surely as $T \to \infty$,
regardless of Player~2's strategy.
\end{blackwelldef}

\begin{blackwellthm}{Blackwell Approachability Theorem (Blackwell, 1956)}{approach}
A closed convex set $\mathcal{S} \subset \mathbb{R}^d$ is approachable by Player~1 if and
only if it is \emph{response-satisfiable}: for every halfspace $H$ containing $\mathcal{S}$,
Player~1 has a mixed strategy $p^* \in \Delta(\mathcal{A})$ such that
$\mathbb{E}_{a \sim p^*}[r(a, b)] \in H$ for all $b \in \mathcal{B}$.

The constructive Blackwell algorithm is:
\begin{enumerate}[noitemsep,label=(\roman*)]
  \item Let $\bar{r}_t$ be the current average payoff and $\pi_t = \Pi_{\mathcal{S}}(\bar{r}_t)$
        its projection onto $\mathcal{S}$.
  \item Play the mixed strategy that minimizes
        $\langle \bar{r}_t - \pi_t,\; \mathbb{E}[r(\cdot, b)] \rangle$ over the worst-case $b$.
\end{enumerate}
This guarantees $d(\bar{r}_T, \mathcal{S}) = O(1/\sqrt{T})$.
\end{blackwellthm}

The theorem generalizes von Neumann's minimax theorem: for $d = 1$ and
$\mathcal{S} = (-\infty, 0]$, approachability reduces exactly to the minimax result.

\subsection{Applications in AI and Machine Learning}

\subsubsection{Equivalence to No-Regret Online Learning}

Perhaps the most far-reaching connection between Blackwell's work and modern AI is the
equivalence established by \citet{abernethy2011blackwell}: any Blackwell approachability
algorithm can be converted into a no-regret algorithm for online linear optimization, and
vice versa. This means the entire family of no-regret algorithms---Multiplicative Weights
Update, Follow-the-Regularized-Leader, Online Mirror Descent---can be understood as special
cases of Blackwell's 1956 framework.
\citet{cesabianchi2006prediction} provide the canonical textbook treatment of prediction with
expert advice; \citet{lattimore2020bandit} unify bandit algorithms under the same framework
and explicitly trace connections to approachability.
No-regret learning is not merely theoretical: it drives convergence of correlated equilibria
in multi-agent systems, underpins adversarial network training, and powers online
recommendation and advertising systems.

\subsubsection{Calibrated Forecasting}

Calibration is a fundamental desideratum for probabilistic AI systems: a model is calibrated
if its stated 90\% confidence events occur 90\% of the time.
\citet{foster1998asymptotic} and \citet{foster1999calibration} showed that the calibration
problem can be reduced to Blackwell approachability---proving that calibrated forecasting is
achievable against any adversarial data-generating process, a much stronger guarantee than
i.i.d.\ results. Modern work on faster recalibration via approachability
\citep{noarov2023faster} has direct applications to calibrating LLMs and probabilistic
classifiers.

\subsubsection{Reinforcement Learning for MDPs}

Blackwell approachability provides an alternative theoretical foundation for RL in Markov
Decision Processes. By constructing auxiliary Blackwell games whose approachable sets
correspond to optimal value functions, one can derive value iteration and Q-learning as
special cases of Blackwell's algorithm. This provides new analytical tools for proving
convergence and regret bounds in RL algorithms.

\subsubsection{Multi-Objective RLHF and LLM Alignment}

Standard RLHF maximizes a single scalar reward derived from human preferences.
However, real human values are multi-dimensional: users have different preferences for safety,
helpfulness, conciseness, tone, and factuality.
\citet{chakraborty2024maxmin} and \citet{xiong2025multiobjective} in the MaxMin-RLHF
framework formulate multi-objective alignment as a vector-payoff game, which is the native
setting for approachability: the target set is the Pareto frontier of human preferences, and
Blackwell's algorithm---project the current reward vector onto the target set, update the
policy toward the best response---supplies both a training recipe and, \emph{under the
assumptions of that formulation}, a convergence argument.

Two qualifications matter, because the guarantee does not transfer intact to deployed systems.
First, Blackwell approachability assumes a convex target set and exact best responses against
vector payoffs. Production alignment optimizes a non-convex neural policy using an
approximate, learned value function, so the classical hypotheses are not satisfied and the
convergence result does not apply directly. Second, this is a research formulation rather than
a description of mainstream practice: widely deployed RLHF pipelines are not implemented as
approachability algorithms. The relationship to them is structural---a shared problem
shape---rather than an identity, and it should be read as such. Establishing
approachability-style guarantees in the non-convex, function-approximation regime remains an
open problem (Section~\ref{sec:open}).

\subsubsection{Fair Online Learning}

\citet{chzhen2021unified} formalized fair online learning as a Blackwell approachability
problem: the agent's decisions must approach a convex set encoding the fairness-accuracy
trade-off frontier. This enables provably fair online learning algorithms with regret
guarantees, applicable to sequential decision systems in hiring, lending, and content
moderation.

%% ════════════════════════════════════════════════════════════════════════════
\section{The Informativeness Theorem}
%% ════════════════════════════════════════════════════════════════════════════

This section states the Informativeness theorem, also known as the comparison of experiments,
and applies it to information design, alignment and safety, and active learning. Where the
previous two theorems concern how to estimate and how to act, this one concerns which data to
gather in the first place: it induces a partial order on information sources that holds
independently of the decision-maker's objective and prior.

\subsection{Theorem Statement and Significance}

\begin{blackwelldef}{Statistical Experiment}{experiment}
A \emph{statistical experiment} (or information structure) $\sigma$ is a pair
$(\Omega, \{P_\theta\}_{\theta \in \Theta})$ where $\Omega$ is a signal space and each
$P_\theta$ is a distribution over signals $\omega \in \Omega$ when the state is $\theta$.
Given two experiments $\sigma$ and $\sigma'$, we say $\sigma$ is \emph{more informative}
than $\sigma'$ (written $\sigma \succeq_B \sigma'$) if the conditions of
Theorem~\ref{thm:informativeness} hold.
\end{blackwelldef}

\begin{blackwellthm}{Blackwell Informativeness Theorem (Blackwell, 1951, 1953)}{informativeness}
For two experiments $\sigma$ and $\sigma'$, the following three conditions are equivalent:
\begin{enumerate}[noitemsep,label=(\roman*)]
  \item \textbf{Garbling:} $\sigma'$ can be obtained from $\sigma$ by applying a stochastic
        kernel (Markov matrix) $K$: $\sigma' = K\sigma$. That is, $\sigma'$ is a
        ``noisy version'' of $\sigma$.
  \item \textbf{Feasibility:} Every decision strategy achievable under $\sigma'$ is also
        achievable under $\sigma$ (i.e.\ $\sigma$ enables a weakly larger set of strategies).
  \item \textbf{Universal preference:} Every Bayesian decision-maker with any prior and any
        loss function weakly prefers $\sigma$ to $\sigma'$, regardless of the decision problem.
\end{enumerate}
When these conditions hold, we say $\sigma$ \emph{Blackwell-dominates} $\sigma'$, written
$\sigma \succeq_B \sigma'$. The relation $\succeq_B$ defines a partial order---the
\emph{Blackwell order}---on the space of experiments.
\end{blackwellthm}

The equivalence of these three conditions---one structural, one strategic, one
decision-theoretic---is the theorem's remarkable depth. It defines a single, objective,
decision-theoretic criterion for when more information is unambiguously better.

\subsection{Applications in AI and Machine Learning}

\subsubsection{Information Design and Mechanism Design}

Information design---determining what information a principal should reveal to an agent to
induce desirable behavior---is a core problem in AI-driven platform economics, recommender
systems, and multi-agent coordination. The canonical treatment is \citet{bergemann2019information},
who provide a comprehensive survey of information design in the Journal of Economic Literature,
establishing the framework in which the designer chooses a signal structure (information policy)
to persuade agents. The Blackwell order provides the natural mathematical language for comparing
information policies: a more informative signal (in the Blackwell sense) allows the agent to
make better decisions for any objective, while a garbled signal is uniformly worse. In
principal-agent models with AI intermediaries, the Blackwell ordering determines the cost of
information asymmetry.

\subsubsection{AI Alignment and Safety}

\citet{alignmentforum2023blackwell} formalizes the Blackwell order as a model of what an AI
system ``knows''---with more informative representations being Blackwell-dominant over less
informative ones. An AI whose world model is a garbling of reality will make suboptimal
decisions for \emph{any} objective, regardless of how well-calibrated its decision procedure.
This provides a principled, objective-independent criterion for evaluating an AI's information
state, connecting foundational statistics to AI safety. A better representation is one that
is less of a garbling of the ground truth.

\subsubsection{Active Learning and Experimental Design}

Bayesian experimental design and active learning---where an AI system chooses which data
points to query to reduce uncertainty most efficiently---are naturally framed in terms of the
Blackwell order. Among candidate experiments, a Blackwell-dominant experiment is
unconditionally preferred: it provides more decision-relevant information regardless of model
and prior. This is particularly relevant in scientific AI applications such as molecular
design, drug discovery, and materials science, where the experimenter's objective may be
partially unknown or evolving.

%% ════════════════════════════════════════════════════════════════════════════
\section{Synthesis: A Unified Information-Theoretic Framework}
%% ════════════════════════════════════════════════════════════════════════════

\subsection{Three Theorems, One Framework}

The three theorems surveyed here---Rao-Blackwell, Approachability, and Informativeness---are
not isolated results. They share a common intellectual core: each is a rigorous statement
about how to extract maximum value from information under uncertainty. Table~\ref{tab:synthesis}
summarizes the conceptual correspondence.

\begin{table}[h!]
\centering
\caption{Unified view of Blackwell's three theorems.}
\label{tab:synthesis}
\renewcommand{\arraystretch}{1.35}
\begin{tabularx}{\textwidth}{l X X}
\toprule
\textbf{Theorem} & \textbf{Central Question} & \textbf{Core AI Principle} \\
\midrule
Rao-Blackwell (1947)
  & How should I \emph{compress} what I know?
  & Condition on sufficient statistics; never discard information relevant to the estimand. \\[4pt]
Approachability (1956)
  & How should I \emph{act} given what I know?
  & In repeated vector-payoff games, project toward the target and play the best response; guarantees $O(1/\sqrt{T})$ regret. \\[4pt]
Informativeness (1951)
  & What data should I \emph{collect}?
  & Prefer Blackwell-dominant information sources; a garbled signal is universally inferior. \\[4pt]
Discounted DP (1965)
  & What is the \emph{long-run optimal} policy?
  & Stationary optimal policies exist and are unique under discounting; value iteration converges. \\
\bottomrule
\end{tabularx}
\end{table}

The Rao-Blackwell theorem addresses \emph{information compression}: the sufficient statistic
is the lossless compression of data for the parameter of interest, and conditioning on it can
only improve estimation. The Approachability theorem addresses \emph{sequential action under
uncertainty}: an agent can guarantee convergence to a target set in vector-payoff space by
systematically responding to information revealed by past play. The Informativeness theorem
addresses \emph{information valuation}: it provides an objective, decision-theoretic criterion
for when one source of information is unconditionally better than another.

\subsection{The AI Triangle}

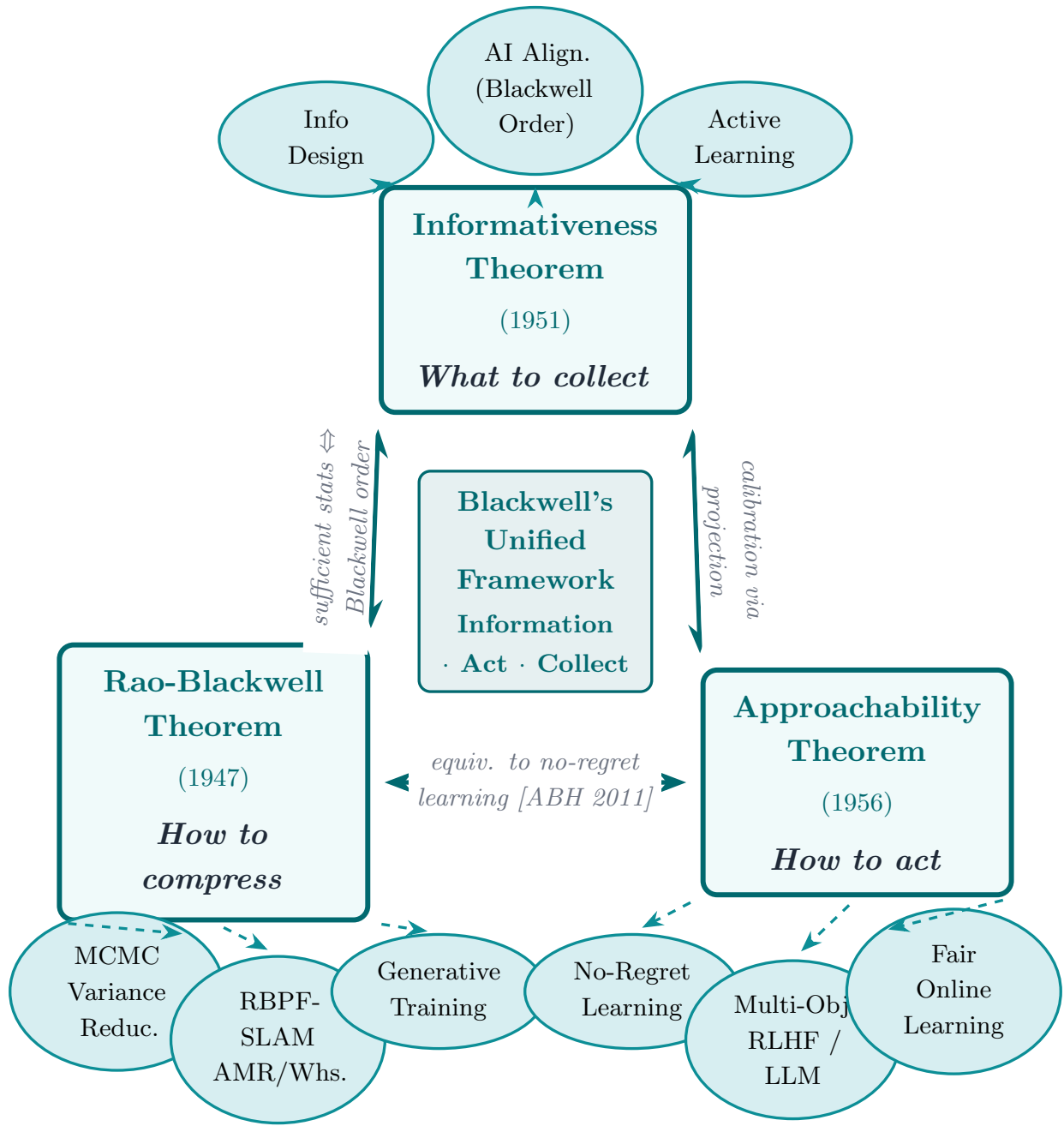
\begin{figure}[H]
\centering
\resizebox{\textwidth}{!}{%
\begin{tikzpicture}[
  >=Stealth,
  %% Theorem vertex nodes
  thmnode/.style={
    rectangle, rounded corners=5pt,
    draw=TealDark, line width=1.8pt,
    fill=TheoremBg,
    text width=3.2cm, align=center,
    inner sep=9pt,
    font=\bfseries\normalsize
  },
  %% AI application nodes
  ainode/.style={
    ellipse,
    draw=TealMid, line width=1pt,
    fill=TealLight!60,
    text width=1.6cm, align=center,
    inner sep=4pt,
    font=\footnotesize
  },
  %% Central box
  centerbox/.style={
    rectangle, rounded corners=4pt,
    fill=TealDark!10, draw=TealDark, line width=1pt,
    text width=2.4cm, align=center,
    inner sep=7pt,
    font=\small\bfseries\color{TealDark}
  },
  %% Arrow styles
  thmedge/.style={TealDark, line width=1.6pt, <->, shorten >=5pt, shorten <=5pt},
  aiedge/.style={TealMid, line width=1pt, dashed, ->, shorten >=4pt, shorten <=4pt},
  elabel/.style={font=\footnotesize\itshape, text=MutedGray,
                 fill=white, inner sep=2pt, midway,
                 text width=3cm, align=center},
]

%% ── Theorem vertices ─────────────────────────────────────────────────────────
%% Triangle: RB bottom-left, APP bottom-right, INFO top-center
\node[thmnode] (RB) at (0, 0) {
  \color{TealDark}Rao-Blackwell\\Theorem\\[2pt]
  {\normalfont\footnotesize (1947)}\\[5pt]
  \color{NavyText}\textit{How to}\\\textit{compress}};

\node[thmnode] (APP) at (8, 0) {
  \color{TealDark}Approachability\\Theorem\\[2pt]
  {\normalfont\footnotesize (1956)}\\[5pt]
  \color{NavyText}\textit{How to act}};

\node[thmnode] (INFO) at (4, 6) {
  \color{TealDark}Informativeness\\Theorem\\[2pt]
  {\normalfont\footnotesize (1951)}\\[5pt]
  \color{NavyText}\textit{What to collect}};

%% ── Central unifying label ───────────────────────────────────────────────────
\node[centerbox] at (4, 2.5) {
  Blackwell's\\Unified\\Framework\\[3pt]
  {\footnotesize Information\\$\cdot$~Act~$\cdot$~Collect}};

%% ── Theorem--theorem edges ───────────────────────────────────────────────────
\draw[thmedge]
  (RB.east) -- (APP.west)
  node[elabel] {equiv. to no-regret learning [ABH 2011]};

\draw[thmedge]
  (RB.north east) -- (INFO.south west)
  node[elabel, sloped, above] {sufficient stats $\Leftrightarrow$ Blackwell order};

\draw[thmedge]
  (APP.north west) -- (INFO.south east)
  node[elabel, sloped, above] {calibration via projection};

%% ── AI application nodes ─────────────────────────────────────────────────────
%% RB children (below-left)
\node[ainode] (MCMC) at (-1.2, -2.6) {MCMC\\Variance\\Reduc.};
\node[ainode] (SLAM) at ( 0.8, -3.2) {RBPF-SLAM\\AMR/Whs.};
\node[ainode] (GEN)  at ( 2.8, -2.6) {Generative\\Training};

%% APP children (below-right)
\node[ainode] (NREG) at (5.2,  -2.6) {No-Regret\\Learning};
\node[ainode] (RLHF) at (7.2,  -3.2) {Multi-Obj.\\RLHF /\\LLM};
\node[ainode] (FAIR) at (9.2,  -2.6) {Fair Online\\Learning};

%% INFO children (above)
\node[ainode] (INFD)  at (1.4, 8.0) {Info\\Design};
\node[ainode] (ALIGN) at (4.0, 8.6) {AI Align.\\(Blackwell\\Order)};
\node[ainode] (ACTV)  at (6.6, 8.0) {Active\\Learning};

%% ── Theorem-to-application edges ─────────────────────────────────────────────
\draw[aiedge] (RB.south west) -- (MCMC.north east);
\draw[aiedge] (RB.south)      -- (SLAM.north);
\draw[aiedge] (RB.south east) -- (GEN.north);

\draw[aiedge] (APP.south west) -- (NREG.north);
\draw[aiedge] (APP.south)      -- (RLHF.north);
\draw[aiedge] (APP.south east) -- (FAIR.north west);

\draw[aiedge] (INFO.north west) -- (INFD.south east);
\draw[aiedge] (INFO.north)      -- (ALIGN.south);
\draw[aiedge] (INFO.north east) -- (ACTV.south west);

\end{tikzpicture}
}% end \resizebox
\caption{The Blackwell AI triangle. The three theorem nodes (rectangles) form the vertices;
  solid double-headed arrows show theoretical equivalences between them;
  dashed arrows connect each theorem to its AI application domains (ellipses).}
\label{fig:triangle}
\end{figure}

Figure~\ref{fig:triangle} shows the relationships visually. The three theorems occupy the
vertices of a triangle, each addressing one leg of the AI triad: \textit{represent}, \textit{act},
and \textit{collect}. The edges between them represent the theoretical connections---the
equivalence between approachability and no-regret learning, the link from
informativeness to sufficient statistics, and the connection from Rao-Blackwellization to
variance reduction in action selection.

\subsection{Temporal Displacement and Mathematical Prescience}

What makes Blackwell's work particularly remarkable from the perspective of AI history is its
\emph{temporal displacement}: results derived in the 1940s and 1950s, without digital
computers in mind, turned out to be exactly the right tools for problems that became
computationally tractable only decades later. The Rao-Blackwell theorem (1947) preceded
practical MCMC by 40 years. The Approachability theorem (1956) preceded its modern online
learning applications by 50 years. The Informativeness theorem (1951) preceded the modern
literature on information design and AI alignment by 60 years.

This pattern---pure mathematics arriving early, waiting, then becoming indispensable---is
not accidental. Blackwell worked at the deepest level of abstraction, asking: what does it
mean to have information? To use information optimally? To compare information sources?
These are not engineering questions; they are philosophical ones with mathematical answers.
The fact that modern AI, at scale, keeps returning to these same questions is testament both
to the depth of Blackwell's insight and to the underlying continuity between classical
statistics and contemporary machine intelligence.

\subsection{Connection to NVIDIA's Blackwell Architecture}

In March 2024, NVIDIA CEO Jensen Huang unveiled the Blackwell GPU architecture at GTC---named
explicitly in honor of David Harold Blackwell. The Blackwell chip contains 208 billion
transistors, manufactured on a 4NP custom TSMC process, and delivers up to 20 petaflops of
AI compute per chip, reducing inference cost by up to $25\times$ relative to the Hopper
architecture \citep{nvidia2024blackwell}.
The naming is symbolically significant: NVIDIA chose to honor a statistician, recognizing
that the statistical frameworks developed in the mid-twentieth century are not merely
historical precursors but active ingredients in modern AI. The variance reduction ideas
embedded in training algorithms, the game-theoretic foundations of alignment methods, and the
information-theoretic criteria for representation quality all trace conceptual lineage to
Blackwell's theorems.

%% ════════════════════════════════════════════════════════════════════════════
\section{Open Problems and Future Directions}
\label{sec:open}
%% ════════════════════════════════════════════════════════════════════════════

Despite the depth of Blackwell's influence on modern AI, several important open problems
remain at the intersection of his theorems and current research frontiers.

\subsubsection*{Blackwell Order for Evaluating LLM Representations}
A fundamental but unresolved question in LLM research is how to compare the quality of
internal representations across model families and architectures. Current evaluation relies
on downstream benchmark performance---a scalar proxy that is necessarily task-specific.
The Blackwell order offers a tantalizing alternative: if one representation can be obtained
from another by a garbling, it is Blackwell-dominated and unconditionally inferior for all
tasks. Developing practical methods to test for Blackwell dominance between LLM representation
spaces---possibly via probing classifiers or information-theoretic estimators---would provide
a task-agnostic benchmark for representation quality.

\subsubsection*{Approachability with Non-Convex Target Sets}
Blackwell's original theorem requires the target set $\mathcal{S}$ to be convex. However,
many realistic alignment objectives---such as Pareto frontiers of reward trade-offs in
multi-objective RLHF---are non-convex. Extending the approachability framework to non-convex
target sets, or characterizing weaker guarantees, is an active area connecting online learning
theory to AI alignment.

\subsubsection*{Rao-Blackwellization for Diffusion Model Training}
Diffusion models train by learning to reverse a stochastic noise process. The training
objective involves Monte Carlo estimates of score functions at each noise level, with
potentially high variance. Whether Rao-Blackwell variance reduction can be systematically
applied to diffusion model training---by identifying sufficient statistics for the denoising
objective---is an open question with significant practical implications.

\subsubsection*{Blackwell's 1965 DP Theorem in Deep RL}
Blackwell's 1965 result on discounted dynamic programming \citep{blackwell1965discounted}---
proving existence and uniqueness of optimal stationary policies in discounted
infinite-horizon MDPs---is a foundational RL result often attributed to Bellman without
acknowledgment of Blackwell's contribution. Understanding how Blackwell's optimality
conditions interact with function approximation error in deep RL is an open problem with
direct implications for the reliability of modern RL systems.

%% ════════════════════════════════════════════════════════════════════════════
\section{Conclusion}
%% ════════════════════════════════════════════════════════════════════════════

This survey has traced three of David Blackwell's principal theoretical contributions---the
Rao-Blackwell theorem, the Approachability theorem, and the Informativeness theorem---through
their influence on modern artificial intelligence. The three connections are not equally tight,
and distinguishing them is part of the contribution. The Rao-Blackwell link is direct and
operational: Rao-Blackwellized particle filters implement the theorem by name and navigate
production warehouse robots, as Section~\ref{sec:rbpf} documents. The Informativeness link is
conceptual but precise: the Blackwell order supplies the decision-theoretic language for
comparing information structures, whether or not a given system cites it. The approachability
link to language-model alignment is the least settled of the three: it is exact for the
vector-payoff formulations of multi-objective RLHF, but mainstream alignment pipelines are not
built as approachability algorithms, and extending the guarantees to non-convex policies with
learned value functions remains open.

What makes Blackwell's work particularly remarkable is its temporal displacement: results
derived without digital computers anticipated problems that became tractable only decades
later. Pure mathematics arrived early, waited, and became indispensable.

Blackwell himself said of mathematics: ``I've always had a strong feeling that I want to
understand whatever I'm working on, not just formally but deeply.'' The field of AI, as it
matures, is increasingly finding that the depth it needs was already there---in the work of
David Blackwell and the generation of statisticians who built the foundations of modern
inference.

%% ════════════════════════════════════════════════════════════════════════════
\appendix
%% ════════════════════════════════════════════════════════════════════════════
\section{Market Context for Indoor Autonomous Mobile Robots}
\label{app:market}
%% ════════════════════════════════════════════════════════════════════════════

The forecasts collected here describe the commercial setting in which RBPF-SLAM
(Section~\ref{sec:rbpf}) is deployed. They are included for context and support no technical
or mathematical claim made elsewhere in this survey. Projections from commercial research
firms differ substantially in scope, methodology, and definition of the addressable market,
and the spread among the estimates below should be read with that in mind rather than
averaged. Growth rates are compound annual growth rates (CAGR) as reported by each source.

\citet{mrf2025indoor} project the indoor robots market to grow from USD\,22.9B (2025) to
USD\,161.3B by 2035 (CAGR $\approx$21.6\%);
\citet{marketsandmarkets2025amr} forecast the global AMR market at USD\,4.56B by 2030
(CAGR $\approx$15.1\%);
\citet{logisticsiq2025warehouse} project warehouse automation at USD\,55B by 2030 with
automated guided vehicles (AGVs) and AMRs as the largest sub-segment ($\approx$30\% CAGR);
\citet{grandview2024mobile} estimate mobile robotics growing from USD\,25.4B (2024) to
USD\,73.7B by 2030 (CAGR $\approx$20.7\%); and
\citet{abi2024mobile} forecast 2.79M mobile robot shipments by 2030 (CAGR $\approx$24.1\%).
Industry data indicates approximately 4.7M warehouse robots installed across 50{,}000+
warehouses by 2026, with a 500\% increase in logistics-robot sales between 2019 and 2025
\citep{sellerscommerce2025}.

The forecasts share a common pattern---double-digit CAGRs attributed to e-commerce growth,
post-pandemic labor shortages, and demand for same-day fulfilment---with indoor mobile robots
and AMRs as the core technology. The consistency of the direction across independent firms is
more informative than any single figure.

%% ════════════════════════════════════════════════════════════════════════════
%% Bibliography
%% ════════════════════════════════════════════════════════════════════════════

\end{document}